\documentclass[%
reprint,
amsmath,amssymb,
aps,
floatfix,
superscriptaddress
]{revtex4-2}
\usepackage{graphicx}
\usepackage{dcolumn}
\usepackage{bm}
\usepackage{booktabs}
\usepackage{siunitx}
\usepackage{hyperref}
\usepackage{multirow}
\usepackage{makecell}
\usepackage{float}
\usepackage{newunicodechar}
\usepackage[utf8]{inputenc}
\DeclareUnicodeCharacter{2248}{\approx}
\newunicodechar{⁻}{\textsuperscript{-}}
\hypersetup{
    colorlinks=true,
    linkcolor=blue,
    citecolor=blue,
    urlcolor=blue
}
\begin{document}
\preprint{arXiv:2501.xxxxx [cs.CV]}
\title{LightHybridNet: A Lightweight Hybrid Deep Learning Framework \\ for Efficient Multi-Label Ophthalmic Disease Classification in Imbalanced Fundus Images}
\author{Hem Raj Dadral}
\thanks{C4054234@hallam.shu.ac.uk}
\affiliation{School of Computing and Digital Technologies, Sheffield Hallam University, Sheffield, United Kingdom}
\author{Najam Ul Hasan}
\thanks{n.ul-hasan@shu.ac.uk}
\thanks{Corresponding author}
\affiliation{School of Computing and Digital Technologies, Sheffield Hallam University, Sheffield, United Kingdom}
\email{n.ul-hasan@shu.ac.uk}
\author{Naveed Ejaz}

\affiliation{School of Computing, Ulster University, Belfast, United Kingdom}
\email{N.Ejaz@ulster.ac.uk}
\author{Muhammad Aslam Jarwar}
\email{a.jarwar@shu.ac.uk}
\affiliation{School of Computing and Digital Technologies, Sheffield Hallam University, Sheffield, United Kingdom}
\date{\today}
\begin{abstract}
Eye diseases are a major global cause of blindness and vision impairment, highlighting the vital need for accurate and early diagnosis to prevent further deterioration. Despite advances in medical imaging, classifying retinal diseases from fundus images remains challenging due to the complexity and subtlety of visual features. This study aims to develop and validate LightHybridNet, a novel lightweight hybrid deep learning framework, to improve the classification of eye diseases from fundus images. The model incorporates advanced architectural features, such as depthwise separable convolutions and compound scaling mechanisms, to enhance diagnostic accuracy and robustness. To validate the proposed model, the ODIR-5K dataset, comprising 8 disease categories, has been used. To enhance image quality, several pre-processing approaches were applied, including histogram equalisation, colour space conversion, and contrast adjustment, resulting in high-resolution final images. Furthermore, image data augmentation was employed to increase the dataset size, thereby enhancing the model's ability to generalise during training. The hybrid architecture helps the model to concentrate on the most relevant regions of the images, while capturing and integrating details at different scales, significantly improving classification performance. To evaluate its effectiveness, the model was compared with various transfer-learning models. The results show that LightHybridNet consistently outperforms the baselines, achieving a weighted F1-score of 91.51\%, a macro AUC of 98.26\%, and a Cohen's Kappa of 90.14\% on the validation set. These results demonstrate superior accuracy, balanced discrimination across classes, and strong agreement despite severe class imbalance. Furthermore, model interpretability was enhanced by employing various explainable AI methods, such as Grad-CAM, to identify and emphasise key features influencing the decision-making process. LightHybridNet represents a promising advancement in automated ocular disease detection, offering improved diagnostic reliability for clinical use.
\begin{description}
\item[Keywords] Multi-label classification, Lightweight Deep Learning, Fundus Imaging, Class Imbalance, MobileNetV3, EfficientNet, Explainable AI.
\end{description}
\end{abstract}
\maketitle
\section{Introduction}
\label{sec:intro}
Vision impairment and blindness represent one of the most significant and persistent global health challenges of the modern era. Sight plays a fundamental role in education, employment, independence, and overall quality of life. When vision is impaired or lost, individuals often become less productive, rely more on caregivers, and participate less socially. According to the International Agency for the Prevention of Blindness (IAPB) Vision Atlas, more than 1.1 billion people worldwide live with vision loss, primarily due to lack of access to basic eye care services~\cite{iapb2026}. Alarmingly, over 90\% of people with vision impairment reside in low- and middle-income countries, where healthcare infrastructure and specialist availability remain limited~\cite{who2019}. Demographic trends, including global population growth, increasing life expectancy, and rapid ageing, are expected to exacerbate this challenge, with projections estimating that the number of people living with vision loss will rise to approximately 1.7 billion by 2050 if current trends persist~\cite{bourne2021}. The burden extends beyond health, encompassing increased mortality risk, reduced educational attainment, and significant economic losses estimated at hundreds of billions of dollars annually in global productivity~\cite{who2019}. These consequences position eye health as a social, economic, and developmental issue closely linked to the United Nations Sustainable Development Goals, particularly those related to health, education, gender equality, and poverty reduction.
A substantial proportion of preventable vision loss is caused by retinal and ocular diseases that can be detected at early stages through clinical screening. Eye diseases encompass a broad category of conditions, including mild to severe impairments like myopia, diabetic retinopathy (DR), glaucoma, age-related macular degeneration (AMD), cataract, hypertensive retinopathy, and pathological myopia~\cite{flaxman2017}. These diseases often progress silently, with patients remaining asymptomatic until irreversible damage occurs, and their risk increases with advancing age~\cite{wong2020}. Fundus photography is one of the most widely used imaging techniques for retinal examination, being non-invasive, relatively low-cost, and capable of capturing high-resolution images of the retina, optic disc, macula, and retinal vasculature~\cite{abramoff2011}. Different diseases manifest through distinct visual patterns: for instance, DR is characterized by microaneurysms, hemorrhages, and exudates; glaucoma presents with optic disc cupping and nerve fibre layer thinning; AMD is associated with drusen deposits and macular changes; while hypertensive retinopathy involves vascular narrowing and abnormal arteriovenous crossings. Despite its diagnostic value, fundus image interpretation remains heavily dependent on trained ophthalmologists, making it time-consuming, subjective, and prone to inter-observer variability.
Recent advancements in artificial intelligence (AI), particularly deep learning, have introduced new opportunities for automated analysis of medical images. In ophthalmology, convolutional neural networks (CNNs) have demonstrated strong performance in detecting retinal diseases from fundus photographs, often achieving accuracy comparable to human experts in controlled settings~\cite{gulshan2016}. AI-based systems can reduce clinician workload, standardise diagnosis, and enable large-scale population screening~\cite{abramoff2011}. However, much existing research prioritises maximum predictive accuracy without considering real-world deployment constraints. Many high-performing models rely on complex architectures, such as deep CNNs, multi-branch networks, and transformers, that contain tens or hundreds of millions of parameters and require high-end GPUs for inference~\cite{resnet2016,densenet2017,vit2021}. This "heavyweight bias" in medical AI highlights the gap between academic innovation and clinical applicability, rendering such models impractical for resource-limited settings such as rural clinics or mobile screening units.
This research is motivated by the need for equitable access to healthcare technologies. For AI-driven ophthalmic screening to have a meaningful global impact, systems must balance diagnostic accuracy with efficiency, portability, and robustness. Lightweight models capable of on-device inference can reduce reliance on cloud infrastructure, preserve patient privacy, and operate in areas with inconsistent internet connectivity, particularly with emerging smartphone-based fundus cameras and portable devices.
Despite promising advances, a significant scalability bottleneck limits the translation of ophthalmic AI to clinical tools. Existing methods typically achieve high accuracy with computationally heavy models unsuitable for resource-constrained environments, or prioritise efficiency but underperform on severely imbalanced multi-label datasets such as ODIR-5K~\cite{odir2019}. The multi-label nature of retinal diagnosis—where a single image may exhibit overlapping conditions—further compounds challenges, with public datasets showing bias toward common diseases and poor minority class performance.
In response, this study addresses the central research question: Can a lightweight hybrid deep learning architecture achieve reliable multi-label retinal disease classification while maintaining low computational cost, enabling deployment in resource-constrained clinical environments? To this end, we introduce LightHybridNet, a novel lightweight hybrid deep learning framework that enhances eye disease classification from fundus images. Unlike prior methods, LightHybridNet incorporates depthwise separable convolutions to focus on diagnostically relevant regions and compound scaling mechanisms to capture fine- and coarse-grained features, handling the subtlety and complexity of ocular pathology~\cite{mobilenetv3,efficientnet}. These enhancements improve the model's ability to handle high variability in fundus images, enabling accurate and robust classification. To address interpretability, we integrate explainable AI methods such as Grad-CAM, which provide visual interpretations of decisions by highlighting key features~\cite{gradcam}. Such transparency fosters trust in clinicians and yields actionable insights for diagnosis and treatment.

This study examines eight primary ocular disease classes, along with a normal condition, from the ODIR-5K dataset, analysing their effects on vision and highlighting the vital importance of early diagnosis and treatment, as illustrated in Figure 1 (representative examples, including frequent comorbid cases).

\textbf{Key contributions include:}

\begin{itemize}
    
\item This study introduces LightHybridNet, a novel deep learning model combining MobileNetV3 and EfficientNet, optimised for fundus image analysis to improve feature encoding and classification accuracy.

\item This study enhances image quality using advanced pre-processing techniques, including histogram equalisation, colour space conversion, and contrast adjustment, thereby ensuring high-resolution images that improve model performance.

\item This study employs advanced lightweight mechanisms to focus on informative regions in retinal images, thereby enhancing the accuracy and interpretability of predictions.

\item This study integrated explainable AI techniques, such as Grad-CAM, to enhance LightHybridNet's interpretability by identifying key features in its decision-making process, thereby improving clinical reliability.

\item LightHybridNet achieves competitive performance, classifying 8 diverse eye diseases with 91.51\% weighted F1-score, paving the way for reliable real-world clinical deployment.
 
\end{itemize}

These advancements extend the current efficiency-adjusted performance in eye disease classification and highlight the effectiveness of lightweight hybrid mechanisms in managing complex medical imaging data.

The organisation of this study is as follows: Section \ref{sec:relatedwork} reviews and details related works and recent advancements in eye disease classification. Section \ref{sec:opt} describes ophthalmic diseases identified through retinal abnormalities in fundus images. Section \ref{sec:methodology} details the methodology, including data collection, preprocessing techniques, and the architecture of the custom lightweight hybrid model. Section \ref{sec:results} provides a comprehensive evaluation of experimental results and model performance. Finally, Section \ref{sec:conclusion} summarises the findings and discusses their broader implications for advancing automated disease diagnosis through medical imaging.
\section{RELATED WORK}
\label{sec:relatedwork}
Eye diseases are a major concern in healthcare due to their potential to cause significant visual impairment and impact overall quality of life. The optimal treatment and management of these conditions necessitated their expeditious identification and precise diagnosis. In this context, machine learning (ML) and deep learning (DL) models emerged as promising approaches to assist healthcare practitioners in accurately diagnosing ocular illnesses \cite{suha2026systematic}. This section evaluated and analysed several studies on the subject, critically synthesising existing research to provide significant insights and context for future advances in automated eye disease identification. Recent studies have made impressive strides in using ML and DL for detecting eye diseases from images such as fundus images and optical coherence tomography (OCT) scans. 

Chowdhury et al. proposed OCCT, an optimised transformer-based model for classifying eye diseases using optical coherence tomography images~\cite{chowdhury2025}. To address class imbalance, the study used GAN-based data augmentation, increasing the dataset to 130,649 images. OCCT, refined through an ablation study, outperformed Vision Transformer (ViT), Swin Transformer, and CNN models, including DenseNet, ResNet, and MobileNetV2, achieving 97.09\% accuracy. Extensive image preprocessing (morphological erosion, median filtering, alpha-beta correction) enhanced quality. The model exhibited high accuracy despite limited training data, making it a robust and scalable AI tool for retinal disease detection.

Shoaib et al. presented the DiaGAN-CNN model, an advanced deep learning framework for diabetic retinopathy (DR) detection~\cite{shoaib2023}. This model combined GAN-based data augmentation with transfer learning, utilising InceptionResNet-v2 and Inception-v3 for feature extraction, to greatly improve classification accuracy. The study validated its method on the Ocular Disease Intelligent Recognition (ODIR) dataset, achieving 98\% accuracy and precision in identifying DR and other eye disorders. The DiaGAN-CNN model surpassed existing approaches by accurately capturing subtle retinal properties, enabling high-speed, high-precision diagnoses.

Ma et al. ~\cite{ma2025} presented the Intelligent Eye Multimodal Interactive Diagnostic System (IOMIDS), an AI-powered chatbot that used textual and image data for eye illness diagnosis and self-triage. In a two-stage cross-sectional research conducted across three medical centres, IOMIDS analysed 50 disorders using 15,640 data entries. Among its multimodal models, the text + smartphone model achieved the highest diagnostic accuracy (internal: 79.6\%, external: 81.1\%), outperforming the others. Multimodal integration enhanced answer quality and reduced disinformation, highlighting the use of AI Chatbots in automated ophthalmic care.

Malik et al. ~\cite{malik2025} introduced dual-branch semi-supervised learning for multi-class retinal disease classification using Colour Fundus Photography (CFP). This approach integrated labelled and unlabelled data through self-training-based semi-supervised learning, improving diagnostic accuracy, especially in data-scarce environments. The authors evaluated six CNN models on the OIH dataset (4217 images); ResNet50 achieved the highest accuracy (93.14\%), with 93\% sensitivity, 98.37\% specificity, and an AUC of 97.75\%. The study provided insight into the potential of semi-supervised learning in clinical decision support systems (CDSS), making AI-powered ophthalmic diagnostics more accessible and effective.

Kansal et al. ~\cite{kansal2022} presented a DL-based system for automated ocular illness categorisation using the ODIR dataset (5,000 fundus images, 8 disease categories). The study employed transfer-learning models (DenseNet201, EfficientNetB3, and InceptionResNetV2) and proposed a two-level feature-selection method that combined Linear Discriminant Analysis (LDA) with neural-network classifiers (DNN, LSTM, and BiLSTM). The BiLSTM classifier achieved 100\% accuracy on training data and over 98\% accuracy on validation data, demonstrating high efficiency and minimal processing complexity. The model provided a scalable, automated approach to clinical decision support, reducing ophthalmologists' workloads and enhancing early disease detection in resource-constrained environments.

Multi-label fundus classification remained challenging due to overlapping disease features and severe class imbalance. Bhati et al. ~\cite{bhati2023} proposed DKCNet, which employed an Inception-ResNetV2 backbone augmented with dilated convolutions and squeeze-and-excitation (SE) blocks, achieving strong performance on ODIR-5K with a macro AUC of 0.9608. However, its high parameter count (~87.7M) limited deployment feasibility.

Alfahdawi et al. ~\cite{alfahdawi2024} introduced Fundus-DeepNet, which incorporated bilateral image fusion, thereby replicating clinical practice by combining information from the left and right eyes. Similarly, He et al. ~\cite{he2021} proposed the Dense Correlation Network (DCNet), which used a backbone CNN to extract features from paired left and right eye colour fundus photographs, followed by a spatial correlation module to capture pixel-wise inter-eye correlations and refine features for patient-level fusion. Evaluated on the ODIR-5K dataset, DCNet achieved a mean AUC of 0.93 and a mean F1-score of 0.913, with significant improvements over baselines that lacked bilateral correlation modelling. This highlighted the value of exploiting inter-eye relationships for multi-label tasks, though DCNet's ResNet-based backbone incurred higher computational costs compared to lightweight alternatives.

Multimodal approaches further extended this paradigm by integrating clinical metadata with image features, demonstrating improved discrimination in complex cases. Recent studies have also explored ultra-wide-field imaging for multi-disease classification, though the computational demands remain high.

Beyond detection, disease staging was critical for prognosis and treatment planning. Grassmann et al. ~\cite{grassmann2018} demonstrated that CNN ensembles could achieve human-level performance in AMD severity prediction on AREDS. Wu et al. ~\cite{wu2022} showed that ConvNeXT, a modern CNN inspired by transformer design principles, outperformed EfficientNetB7 in AMD staging while maintaining superior efficiency. Attention-based EfficientNet variants were also applied to glaucoma staging with promising results.

Recent trends focused on attention mechanisms, hybrid designs, and data-centric optimisation. ConvNeXT utilised large-kernel convolutions and depthwise separability to capture global context efficiently~\cite{wu2022}. EfficientNetV2 incorporated compound scaling and channel attention to enhance feature learning. Data-centric strategies such as CLAHE preprocessing, aggressive data augmentation, and cost-sensitive learning proved essential for handling dataset imbalance in ODIR-5K.

Heavyweight models were characterised by deep, complex architectures with tens of millions of parameters and high requirements for floating-point operations. These models often employed multi-branch designs, dense connectivity, attention mechanisms, or ensemble-style feature fusion to maximise representational capacity. Examples included Inception-ResNet variants, DenseNet-based hybrids, and large EfficientNet or transformer-based networks. While heavyweight architectures frequently achieved high performance in controlled experimental settings, they demanded substantial computational resources, extended training time, and high memory consumption, limiting their deployment in clinical environments with constrained hardware~\cite{bhati2023}.

In contrast, lightweight models were designed with efficiency as a primary objective. They typically employed depthwise separable convolutions, inverted residual blocks, parameter sharing, and streamlined feature-extraction pipelines to minimise model size and computation while incurring minimal performance loss. Architectures such as MobileNetV3, EfficientNet-B0/B1, and MobileViT exemplified this category, achieving competitive accuracy with an order-of-magnitude fewer parameters and FLOPs~\cite{mobilenetv3}\cite{efficientnet}.

This distinction was crucial in ophthalmic screening applications. Fundus-based diagnostic systems were increasingly deployed on portable cameras, mobile devices, and edge-computing platforms, particularly in low- and middle-income regions where access to ophthalmologists was limited~\cite{who2019}. In such contexts, heavyweight models were often impractical due to latency, energy consumption, and infrastructure requirements. Lightweight architectures, by contrast, enabled real-time inference, lower power usage, and scalable tele-ophthalmology solutions.

However, purely lightweight models sometimes struggled to capture the complex, multi-scale pathological patterns present in multi-label datasets such as ODIR-5K, especially under severe class imbalance. This motivated recent research into lightweight hybrid architectures, which combined complementary efficient backbones to enhance representational capacity while maintaining deployability~\cite{wu2022}.

Despite their high accuracy, DL models were often criticised as “black boxes,” which limited clinical trust and adoption. XAI techniques such as Grad-CAM generated heatmaps that highlighted diagnostically relevant retinal regions, aligning predictions with ophthalmological landmarks~\cite{gradcam}\cite{bhati2023}. Model-agnostic methods, including LIME and SHAP, further quantified feature importance and decision contributions.

Interpretability was particularly critical for lightweight models deployed in real-world screening programs, where transparency, accountability, and clinician trust were essential.

Heavyweight models achieved high diagnostic performance but were impractical for real-world deployment due to computational demands~\cite{bhati2023}. Lightweight models were efficient but often underperformed on complex, imbalanced, multi-label datasets~\cite{mobilenetv3}.

Key gaps included the lack of unified frameworks that simultaneously addressed multi-disease classification, severity staging, and model interpretability. Furthermore, XAI remained underutilised in lightweight architectures.
\section{OPHTHALMIC DISEASES IDENTIFIED THROUGH RETINAL ABNORMALITIES IN FUNDUS IMAGES}
\label{sec:opt}
\begin{figure*}
    \centering
    \includegraphics[width=1\linewidth]{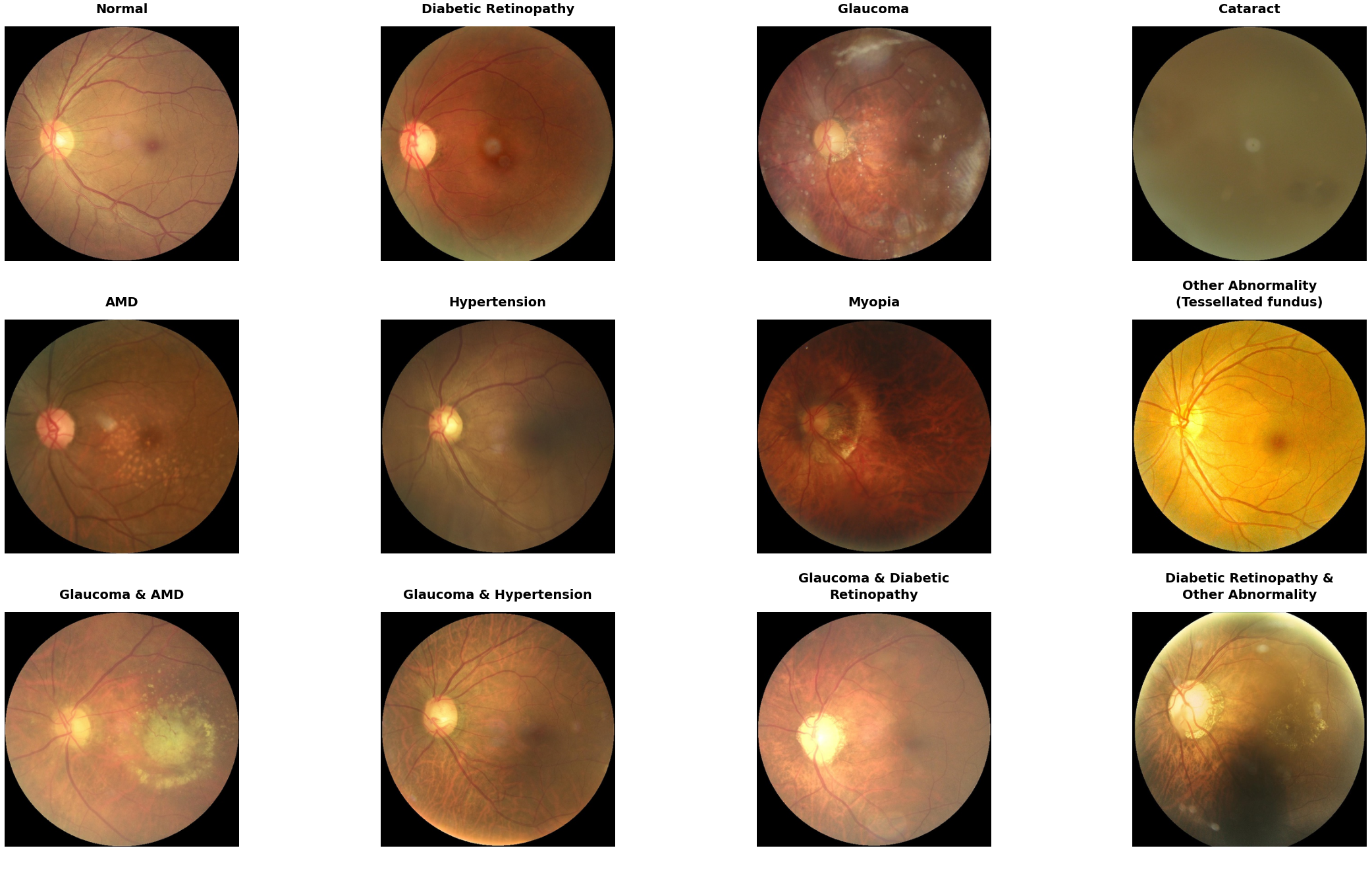}
    \caption{Representative fundus images from the ODIR-5K dataset illustrating the eight primary ocular disease classes (Normal, Diabetic Retinopathy, Glaucoma, Cataract, AMD, Hypertension, Pathological Myopia, Other abnormalities) along with selected examples of common comorbid presentations.}
    \label{fig:Fig1_rep}
\end{figure*}
Figure \ref{fig:Fig1_rep} shows fundus images belonging to different disease classes from the ODIR-5K dataset. Glaucoma is an eye condition that affects the optic nerve, whose strength is crucial for good vision. This harm is typically caused by an unusually high pressure in the eye. Glaucoma can be distinguished by noticing changes in the proportion of the optic disc cup and neuroretinal edge surface region known as the cup-to-disc ratio, along with nerve fibre layer thinning and optic disc haemorrhages. Diabetic retinopathy is a complication of diabetes that affects the tissues inside the retina. Diabetic retinopathy may not cause any adverse signs, and patients may have only minor vision issues. However, it may lead to visual impairment if not treated at an earlier stage. The macula is responsible for clear central vision, and vision becomes distorted if fluid accumulates in it. Age-related Macular Degeneration (AMD) can be distinguished by the presence of neovascularisation (growth of new blood vessels) and geographic atrophy (death of retinal cells), along with drusen deposits and pigmentary changes. The cause of cataracts includes the development of patches that make vision difficult. The presence of cataracts can be observed when the optic disc, fovea, lens, and other parts of the eye become hazy with diffuse clouding. Hypertension is a silent illness. This disease changes the biological characteristics of veins, such as length and thickness, resulting in arteriolar narrowing, arteriovenous nicking, flame haemorrhages, and cardiovascular complications over the long run. Myopia, particularly pathological myopia, is an eye condition that can cause significant vision loss due to thinning of the corneal tissue, peripheral retinal degeneration, and changes in eye colour. It fundamentally modifies the visualisation of any object from a certain distance, making it appear blurred with tessellated fundus and peripapillary atrophy. The fundus images can also be used to detect other types of anomalies, such as pigment epithelium proliferation, epiretinal membrane, tessellated fundus, vitreous macular degeneration, and mixed pathologies.
\section{METHODOLOGY}
\label{sec:methodology}
\subsection{ODIR-5K Dataset}
The ODIR-5K dataset used in this work is available online through a grand challenge hosted by Peking University~\cite{odir2019}. The dataset contains around 5000 organised fundus images of the left and right eyes of patients. These fundus images were captured using various fundus cameras, such as Kowa, Zeiss, and Canon, at different resolutions. Disease diagnostic keywords were assigned to these images by eye specialists. Based on these diagnostic keywords, the disease classification labels are assigned to each pair of fundus images. This visual pathology dataset is unique among publicly accessible datasets, as it contains colour fundus images of both the left and right eyes of a patient, with single or multiple abnormalities per image. The dataset contains a common target label for the pair of eye images. The images are grouped into eight disease classes: Normal, Diabetes, Glaucoma, Cataract, AMD, Hypertension, Myopia, and Others. Patients’ age and gender are also included. Another challenging aspect of the ODIR-5K dataset is that images from the other class contain lesions associated with 12 different ophthalmic diseases. It is not easy to learn appropriate features in such a case. Also, the dataset is highly imbalanced, given the number of images per class. These issues negatively affect accuracy and reduce the performance of the trained classification models. Although the ODIR-5K dataset is more applicable to real-life clinical situations, as the images are captured with different cameras under varying illumination conditions, this poses a significant challenge for any disease identification model.

To ensure that only diagnostically reliable images were used, a quality-filtering step was applied after loading the metadata and mapping image paths. The filtering criteria were explicitly defined as follows:
\begin{itemize}
    \item Missing or corrupted image files (path does not exist or file cannot be opened).
    \item Presence of non-diagnostic quality-issue keywords in the original dataset metadata, specifically: ``low-quality image'', ``lens dust'', ``optic disk photographically invisible'', or ``image offset''.
\end{itemize}
A total of \textbf{324 images} ($\approx$3.2\% of the initial set) were removed due to these conditions, resulting in a final dataset of \textbf{9,676 valid fundus images} (approximately 4,838 left-eye and 4,838 right-eye images after pairing verification). This strict filtering minimises the risk of the model learning from imaging artefacts or noise.

Regarding label assignment: The ODIR-5K dataset provides patient-level multi-label annotations (one 8-dimensional binary vector per patient), derived from diagnostic keywords associated with both left and right eyes. In this work, left and right fundus images are treated as separate, independent inputs to the CNN. To enable per-image training while remaining consistent with the dataset's design, the full patient-level label vector is assigned identically to both the left-eye and right-eye images of the same patient. This approach is widely adopted in ODIR-5K literature because diseases can be bilateral (affecting both eyes) or unilateral (affecting one eye), and the patient-level label effectively represents the logical OR of any eye-specific findings. Thus, if a patient has diabetic retinopathy in one or both eyes, both images receive a positive label for Diabetes (D=1). This per-image label propagation allows straightforward convolutional processing while preserving approximate clinical consistency.
\subsection{Pre-processing and Image Enhancement}
The heterogeneous nature of the ODIR-5K dataset, stemming from various camera models and illumination conditions, necessitates rigorous pre-processing. Initially, a field-of-view (FOV) aware cropping operation is performed by identifying the first non-black pixels to define the image mask borders, effectively removing redundant background information. Images are then resized to a uniform $224 \times 224$ pixel resolution, a standard input dimension for the MobileNetV3 and EfficientNet backbones ~\cite{mobilenetv3,efficientnet}.

A critical step in our pipeline is the application of Contrast Limited Adaptive Histogram Equalisation (CLAHE) ~\cite{clahe,arulraj2025}. Unlike global histogram equalisation, which can over-amplify noise in homogeneous regions of fundus images, CLAHE operates on localised $8 \times 8$ tiles. To prevent the ``noise-over-enhancement'' common in medical imaging, we implement a clip limit ($C_L = 2.0$) that truncates the histogram bins. The maximum height of any histogram bin $M$ is defined as:
\begin{equation}
M = C_L \times \frac{N_{\mathrm{pixels}}}{L}
\end{equation}
where $N_{\mathrm{pixels}}$ represents the number of pixels per tile and $L$ denotes the number of grey levels (typically 256). Any values exceeding $M$ are clipped and uniformly redistributed across all bins before calculating the Cumulative Distribution Function (CDF) for intensity mapping. This technique is particularly effective for enhancing the visibility of fine vascular structures and subtle lesions such as microaneurysms and drusen while maintaining anatomical fidelity.

Finally, all images undergo ImageNet normalisation (mean $[0.485, 0.456, 0.406]$ and standard deviation $[0.229, 0.224, 0.225]$) to standardise the input distribution for the pre-trained backbones.
\subsection{Data Augmentation Strategy}
To improve model generalisation, simulate real-world acquisition variability, and mitigate overfitting—particularly to replicated minority-class samples—we applied an extensive on-the-fly augmentation pipeline exclusively during training, using the Albumentations library ~\cite{albumentations}.

\begin{figure}[H] % [H] forces the figure to stay exactly here (after the paragraph)
    \centering
    \includegraphics[width=1\linewidth]{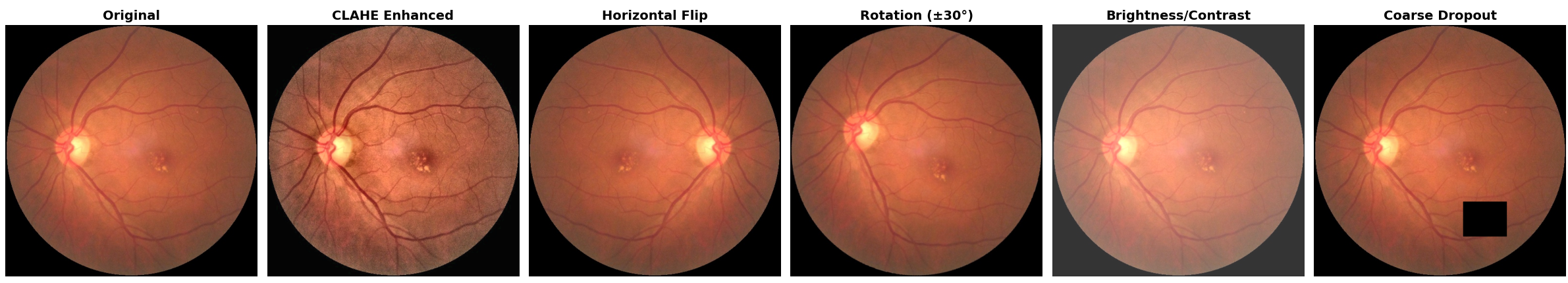}
    \caption{Examples of the applied data augmentation transformations on sample fundus images from the ODIR-5K dataset (original image on the left, followed by augmented variants demonstrating flips, rotations, brightness/contrast changes, CLAHE enhancement, shift-scale-rotate, and coarse dropout).}
    \label{fig:augmentation_pipeline}
\end{figure}

The training augmentation pipeline consists of the following stochastic transformations:
\begin{itemize}
    \item Geometric invariance: Horizontal flip ($p=0.5$), vertical flip ($p=0.3$), and random rotation up to $\pm 30^\circ$ ($p=0.5$) to account for variations in camera orientation and patient positioning.
    \item Illumination variability: Random brightness and contrast adjustments ($\pm 0.2$, $p=0.5$) to simulate differences across imaging devices and clinical environments.
    \item Contrast enhancement: CLAHE with clip limit $C_L=2.0$ ($p=0.5$) to further improve vessel and lesion visibility without excessive noise amplification.
    \item Mild geometric distortions: Shift-scale-rotate (shift limit $=0.1$, scale limit $=0.1$, rotate limit $=15^\circ$, $p=0.6$) to mimic varying patient-to-camera distances.
    \item Regularisation: Coarse dropout (max holes $=8$, max height/width $=16$ pixels, $p=0.5$) to randomly mask contiguous regions, encouraging the model to rely on global context rather than localised artefacts ~\cite{cutout}.
\end{itemize}
As illustrated in Figure~\ref{fig:augmentation_pipeline}, these transformations generate clinically plausible variations while preserving anatomical integrity. Validation and test images undergo only deterministic pre-processing (cropping, resizing, and normalisation) to ensure unbiased evaluation under realistic conditions. Prior studies on retinal benchmarks, including ODIR-5K, have shown that aggressive augmentation strategies of this nature can yield 5--10\% improvements in AUC.
\subsection{Class Balancing}
The left and right fundus images are treated individually as input to the CNN architecture in this work. Based on the disease description for a particular eye part, a disease label is assigned to each image. A fundus image having multiple diseases is treated separately with individual disease class labels. The images with artefacts like “low-quality image”, “optical disk photographically invisible,” “lens dust,” and “image offset” are removed from the final dataset to reduce the false recognition rate. After separating left and right eye images with their corresponding ground truth, the class-wise distribution is shown in Table I. It is observed that the normal, diabetes, and other classes have significantly more images than the other disease classes.

Replication-based oversampling is used in this work to address the class-balance issue without introducing synthetic artefacts ~\cite{shorten2019}. Minority-class samples are duplicated to approximately match the majority-class count, thereby expanding the training set while preserving anatomical fidelity. Aggressive on-the-fly augmentation (flips, rotations, brightness/contrast, CLAHE, coarse dropout) prevents overfitting to duplicates. We chose replication-based oversampling over synthetic methods such as SMOTE or GANs to avoid generating anatomically unrealistic samples that could mislead the model in medical imaging contexts. However, we acknowledge that duplicating minority samples may introduce some memorisation bias, which can be mitigated through aggressive data augmentation.
\begin{table}
\centering
\caption{ODIR-5K DATASET CLASS STATISTICS IN ORIGINAL AND AFTER REPLICATION-BASED OVERSAMPLING OPERATION}
\footnotesize
\setlength{\tabcolsep}{4.2pt}
\renewcommand{\arraystretch}{1.35}
\begin{ruledtabular}
\begin{tabular}{lcc}
Class (Label) & Original Samples & After Oversampling \\
\hline
Normal (N)        & $\sim$3,327 & $\sim$3,327 \\
Diabetes (D)      & $\sim$3,300 & $\sim$3,300 \\
Glaucoma (G)      & $\sim$500   & $\sim$3,300 \\
Cataract (C)      & $\sim$500   & $\sim$3,300 \\
AMD (A)           & $\sim$400   & $\sim$3,300 \\
Hypertension (H)  & $\sim$138   & $\sim$3,300 \\
Myopia (M)        & $\sim$400   & $\sim$3,300 \\
Others (O)        & $\sim$2,500 & $\sim$2,500 \\
\end{tabular}
\end{ruledtabular}
\label{tab:odir_class_distribution}
\end{table}

\subsection{Overview of the Proposed LightHybridNet Architecture}
\begin{figure*}
    \centering
    \includegraphics[width=1\linewidth]{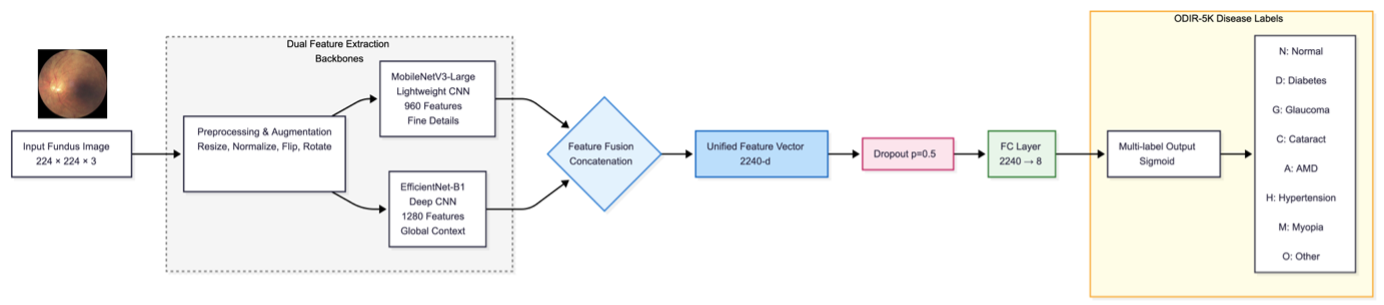}
    \caption{Detailed architecture of the LightHybridNet framework, illustrating the parallel fusion of MobileNetV3-Large and EfficientNet-B1 backbones}
    \label{fig:placeholder}
\end{figure*}
As shown in Fig. 3, LightHybridNet comprises parallel MobileNetV3-Large and EfficientNet-B1 backbones ~\cite{mobilenetv3,efficientnet}, global average pooling, feature concatenation, dropout regularisation, and a lightweight fully connected sigmoid head. The backbone networks extract complementary features: MobileNetV3 excels at local texture and fine-grained lesion detection via depthwise separable convolutions and inverted residuals, while EfficientNet-B1 captures multi-scale global context through compound scaling and MBConv blocks. The resulting 1280-dimensional features from each backbone are concatenated to form a 2560-dimensional representation, followed by dropout (p=0.5) and mapping to 8 sigmoid outputs for multi-label classification. This late fusion strategy preserves independent hierarchies while achieving only 10.74M parameters and 0.853 GFLOPs. Late feature fusion was selected over early fusion to allow each backbone to independently develop specialised representations before integration, reducing computational overhead during training and inference. Attention mechanisms were excluded to maintain low parameter count and FLOPs, prioritising edge deployment feasibility over marginal accuracy gains.

\subsubsection{MobileNetV3-Large (mobilenetv3\_large\_100.ra\_in1k)}
The MobileNetV3-Large backbone is instantiated using \texttt{timm.create\_model('mobilenetv3\_large\_100.ra\_in1k', pretrained=True, num\_classes=0)} ~\cite{timm}. This model is a variant of MobileNetV3, a lightweight convolutional neural network architecture designed for efficient mobile vision applications. The 'large' denotes the larger configuration with a width multiplier of 1.0 (100\% width), optimised for higher accuracy while maintaining low computational cost. The 'ra' suffix indicates training with RandAugment, a data augmentation technique that improves generalisation. 'in1k' refers to pretraining on the ImageNet-1k dataset, which contains 1,000 classes. By setting \texttt{num\_classes=0}, the final classification head is removed, allowing the model to function as a feature extractor and output 1280-dimensional embeddings. MobileNetV3-Large is particularly effective at capturing local textures and fine-grained details in fundus images, thanks to its depthwise separable convolutions, inverted residuals, and squeeze-and-excitation blocks, making it well-suited for detecting subtle lesions such as microaneurysms in diabetic retinopathy.

\subsubsection{EfficientNet-B1 (efficientnet\_b1)}
The EfficientNet-B1 backbone is created via \texttt{timm.create\_model('efficientnet\_b1', pretrained=True, num\_classes=0)}. EfficientNet-B1 is part of the EfficientNet family, which employs compound scaling to balance network depth, width, and resolution for optimal performance-efficiency trade-off. The 'B1' variant is a baseline model with moderate size, pretrained on ImageNet-1k ('in1k' implied). Setting \texttt{num\_classes=0} configures it as a feature extractor, producing 1280-dimensional features. EfficientNet-B1 excels at multi-scale feature capture through its MBConv blocks (mobile inverted bottleneck convolutions) with squeeze-and-excitation optimisation and swish activations. This makes it adept at modelling global pathological structures in retinal images, such as optic disc cupping in glaucoma or macular changes in AMD ~\cite{efficientnet}. The fine-tuning from the original TensorFlow weights ensures compatibility with PyTorch while preserving high transfer-learning efficacy.

\subsection{Evaluation Metrics}
The model was trained using Binary Cross-Entropy with Logits Loss (BCEWithLogitsLoss), which is well-suited for multi-label classification as it independently optimises each label ~\cite{pytorch}:
\begin{equation}
\begin{split}
\mathcal{L}(\mathbf{y}, \hat{\mathbf{y}}) = {} & -\frac{1}{N L} \sum_{i=1}^{N} \sum_{j=1}^{L} \\
& \left[ y_{i,j} \cdot \log(\sigma(\hat{y}_{i,j})) + (1 - y_{i,j}) \cdot \log(1 - \sigma(\hat{y}_{i,j})) \right]
\end{split}
\label{eq:bce}
\end{equation}
where $N$ is the number of samples, $L=8$ is the number of labels, $y_{i,j} \in \{0,1\}$ is the ground-truth label, $\hat{y}_{i,j}$ is the logit output, and $\sigma(\cdot)$ denotes the sigmoid function.
Performance was evaluated using the following metrics, selected for their robustness in multi-label and imbalanced settings ~\cite{sokolova2009,powers2020}:
\begin{enumerate}
    \item \textbf{Macro-averaged AUC.} To assess discriminative ability independently of decision thresholds, we computed the macro-averaged Area Under the ROC Curve (AUC):
    \begin{equation}
    \text{Macro AUC} = \frac{1}{L} \sum_{j=1}^{L} \text{AUC}_j
    \label{eq:macro_auc}
    \end{equation}
    where $\text{AUC}_j$ is obtained via one-vs-rest binarisation for class $j$. This metric treats all classes equally, emphasising performance on rare pathologies.
    \item \textbf{Hamming Score.} As traditional accuracy is inappropriate for multi-label tasks, we report the Hamming Score (label-wise accuracy):
    \begin{equation}
    \text{Hamming Score} = \frac{1}{N L} \sum_{i=1}^{N} \sum_{j=1}^{L} \mathbb{I}( \hat{y}_{i,j} = y_{i,j} )
    \label{eq:hamming}
    \end{equation}
    where $\mathbb{I}(\cdot)$ is the indicator function and $\hat{y}_{i,j}$ is the predicted binary label (thresholded at 0.5).
    \item \textbf{Weighted Precision and Recall.} To account for severe class imbalance, weighted variants were used:
    \begin{align}
    P_j &= \frac{\text{TP}_j}{\text{TP}_j + \text{FP}_j} \tag{6} \\
    \text{Weighted Precision} &= \sum_{j=1}^{L} w_j \cdot P_j \tag{7} \\
    R_j &= \frac{\text{TP}_j}{\text{TP}_j + \text{FN}_j} \tag{8} \\
    \text{Weighted Recall} &= \sum_{j=1}^{L} w_j \cdot R_j \tag{9}
    \end{align}
    where $w_j$ is the support proportion of class $j$ in the dataset.
    \item \textbf{Weighted F1-Score.} The primary optimisation target and reporting metric was the weighted F1-score, the harmonic mean of weighted precision and recall:
    \begin{align}
    F1_j &= 2 \cdot \frac{P_j \cdot R_j}{P_j + R_j} \tag{10} \\
    F1_{\text{weighted}} &= \sum_{j=1}^{L} w_j \cdot F1_j \tag{11}
    \end{align}
    \textbf{Risk Prediction Scoring.}  
To support clinical triage and referral prioritisation, a composite Risk Score $S$ is computed from the model's per-class probabilities $p_j = \sigma(\hat{y}_j)$ using clinically motivated severity weights $\Omega_j$:

\begin{equation}
S = \sum_{j=1}^{8} p_j \cdot \Omega_j
\label{eq:risk_score}
\end{equation}

The weights $\Omega_j$ are assigned based on estimated risk of vision loss and urgency of intervention (higher = greater clinical priority)~\cite{who2019,flaxman2017,wong2020}:

\begin{itemize}
    \item Glaucoma (G): $\Omega_G = 3.0$ (high irreversible optic nerve damage risk)
    \item AMD (A): $\Omega_A = 2.5$ (leading cause of severe central vision loss in the elderly)
    \item Diabetic Retinopathy (D): $\Omega_D = 2.0$ (progressive, treatable if detected early)
    \item Cataract (C): $\Omega_C = 1.8$ (highly treatable, common cause of reversible impairment)
    \item Hypertension (H): $\Omega_H = 1.5$ (systemic risk indicator, often subtle)
    \item Pathological Myopia (M): $\Omega_M = 1.5$ (progressive retinal degeneration)
    \item Others (O): $\Omega_O = 1.0$ (heterogeneous, variable severity)
    \item Normal (N): $\Omega_N = 0.0$ (no contribution to risk)
\end{itemize}

Higher $S$ values indicate greater overall ocular risk, enabling screening programs to prioritise patients for specialist referral. This simple yet interpretable scoring aligns model outputs with clinical decision-making without requiring additional training.
\end{enumerate}

\section{RESULTS}
\label{sec:results}
\subsection{Experimental Setup}
All experiments are conducted on the preprocessed ODIR-5K dataset with a patient-level stratified split (80\% training, 20\% validation), ensuring that left- and right-eye images from the same patient are not separated between training and validation sets to prevent data leakage ~ \ cite {odir2019}. After quality filtering and left/right eye separation, the training set contains approximately 9,968 images. Replication-based oversampling expands minority classes to match the majority class count ($\sim$3,300), resulting in $\sim$24,933 training samples. Extensive on-the-fly augmentation (horizontal/vertical flips, rotations $\pm 30^\circ$, brightness/contrast $\pm 0.2$, CLAHE clip limit 2.0, coarse dropout) is applied to prevent overfitting to duplicated samples ~\cite{shorten2019,cutout}.

An 80/20 patient-level stratified split was chosen to balance computational efficiency, stable evaluation of the minority class, and sufficient data for model development on the moderately sized, severely imbalanced ODIR-5K dataset. While 5-fold cross-validation or a separate held-out test set could provide even more robust estimates, the current configuration ensures no leakage between fellow eyes and allocates adequate validation samples ($\sim$20\%, $\sim$2,500 images) for hyperparameter tuning and performance monitoring. This split is widely adopted in prior ODIR-5K studies due to the dataset's size and imbalance, where smaller folds risk unstable metrics for rare classes. All final reported results (weighted F1-score, macro AUC, Cohen's Kappa) are computed on the held-out validation portion to provide a realistic estimate of generalisation performance. Future work may explore full k-fold CV or external test sets for additional validation.

LightHybridNet is trained for 25 epochs using Binary Cross-Entropy with Logits loss, AdamW optimiser (learning rate $3 \times 10^{-4}$, weight decay $1 \times 10^{-2}$) ~\cite{adamw}, and OneCycleLR scheduler ~\cite{onecycle}. Mixed-precision training (FP16) is used on NVIDIA GPUs ~\cite{pytorch}. Evaluation metrics include weighted F1-score, macro AUC, Cohen's Kappa ~\cite{powers2020,kappa1968}, per-class precision/recall/F1, and efficiency measures (parameters, GFLOPs, inference latency).

Ablation studies compare imbalance strategies and backbone combinations. The proposed model is benchmarked against heavyweight baselines (DKCNet~\cite{bhati2023}, DCNet~\cite{he2021}) and lightweight alternatives.
\subsection{Impact of Class Imbalance Mitigation}
Severe class imbalance in ODIR-5K negatively affects minority class performance~\cite{bhati2023}. Three strategies were evaluated:
\begin{table}[H]
\centering
\caption{PERFORMANCE COMPARISON OF CLASS IMBALANCE MITIGATION STRATEGIES ON LIGHTHYBRIDNET}
\footnotesize
\setlength{\tabcolsep}{4.0pt}
\renewcommand{\arraystretch}{1.35}
\begin{ruledtabular}
\begin{tabular}{lcccc}
Strategy & Samples & F1-Score & AUC & Kappa \\
\hline
\makecell[l]{Weighted Random\\Sampler (Baseline)} 
& 9,968 & 0.8813 & 0.9749 & 0.8616 \\
\makecell[l]{Replication-Based\\Undersampling} 
& $\sim$5,328 & 0.6858 & 0.9213 & 0.6361 \\
\makecell[l]{Replication-Based\\Oversampling\\\textbf{(Proposed)}} 
& $\sim$24,933 & \textbf{0.9151} & \textbf{0.9826} & \textbf{0.9014} \\
\end{tabular}
\end{ruledtabular}
\label{tab:performance_comparison}
\end{table}

\begin{figure}[H] % <--- Changed to [H] (definitely here)
    \centering
    \includegraphics[width=0.95\linewidth]{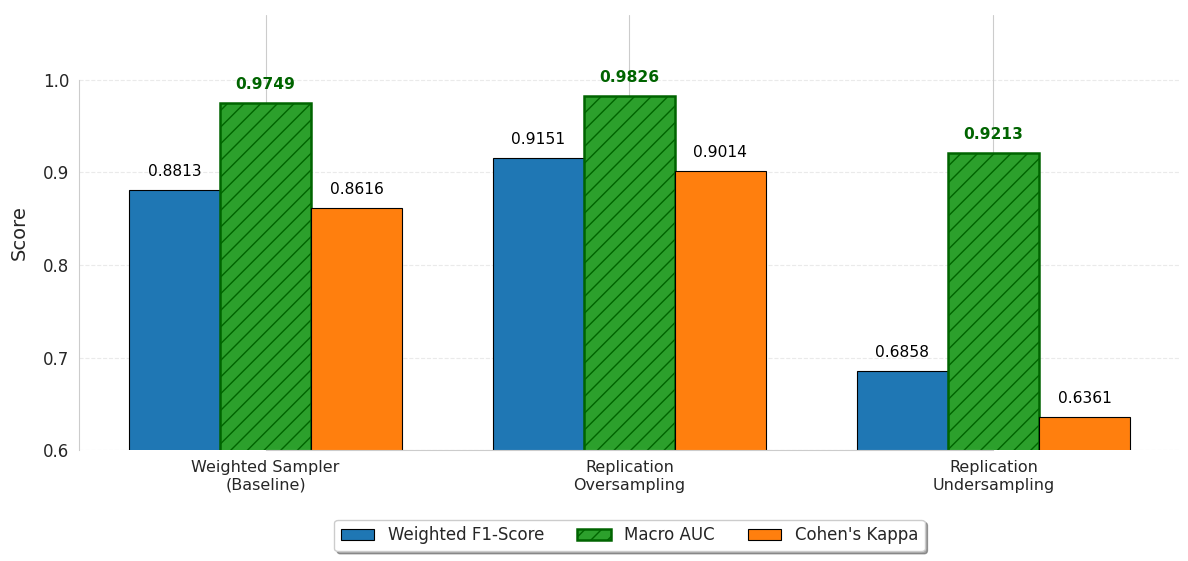}
    \caption{Visualisation of performance comparison across class imbalance mitigation strategies (Table~\ref{tab:performance_comparison}).}
    \label{fig:performance_comparison}
\end{figure}
Replication-based oversampling yields substantial gains (+3.38\% weighted F1, +0.76\% Kappa over baseline), confirming that preserving majority class volume while increasing minority exposure—combined with aggressive augmentation—is critical for learning robust representations of both normal anatomy and rare pathologies ~\cite{shorten2019}.

\subsection{Training Progression}
Training loss decreased smoothly from 0.3590 (Epoch~1) to 0.0043 (Epoch~25). Peak validation performance occurred around Epoch~22--24.

\begin{table}[H]
\centering
\caption{VALIDATION PERFORMANCE PROGRESSION ACROSS KEY EPOCHS}
\footnotesize
\setlength{\tabcolsep}{4.0pt}
\renewcommand{\arraystretch}{1.35}
\begin{ruledtabular}
\begin{tabular}{lccccc}
Epoch & Train Loss & Weighted F1 & Macro AUC & Kappa \\
\hline
1  & 0.3590 & 0.4054 & 0.8401 & 0.3430 \\
5  & 0.1058 & 0.6884 & 0.9286 & 0.6411 \\
10 & 0.0433 & 0.8381 & 0.9677 & 0.8141 \\
15 & 0.0181 & 0.8920 & 0.9760 & 0.8742 \\
20 & 0.0062 & 0.9076 & 0.9802 & 0.8926 \\
25 & 0.0043 & 0.9151 & 0.9826 & 0.9014 \\
\end{tabular}
\end{ruledtabular}
\label{tab:validation_progression}
\end{table}

Stable convergence without overfitting validates the efficacy of replication-based oversampling and augmentation.
\subsection{Comparative performance and efficiency metrics (ablation and benchmarks)}
LightHybridNet (with replication-based oversampling) achieved 10.74M parameters, 0.853 GFLOPs, weighted F1-score 0.9151, macro AUC 0.9826, and Cohen's Kappa 0.9014.

An ablation study was conducted to isolate the contributions of the hybrid architecture and oversampling. Standalone MobileNetV3-Large and EfficientNet-B1 backbones (trained without oversampling) provide strong baselines but are outperformed by the full hybrid model. Hybrid fusion alone yields modest gains, while oversampling delivers the largest improvement (+3.38\% weighted F1 over hybrid without oversampling), confirming that balanced data exposure is critical for realising complementary backbone strengths.

\begin{table}[H]
\centering
\caption{COMPARATIVE PERFORMANCE AND EFFICIENCY METRICS (ABLATION AND BENCHMARKS)}
\footnotesize
\setlength{\tabcolsep}{3.2pt}
\begin{ruledtabular}
\renewcommand{\arraystretch}{1.35}
\begin{tabular}{lccccc}
Model & Params (M) & GFLOPs & F1-Score & AUC & Kappa \\
\hline
MobileNetV3          & 4.21  & 0.237  & 89.54 & 97.36 & 87.87 \\
EfficientNet-B1      & 6.52  & 0.615  & 90.96 & 97.82 & 89.46 \\

EfficientNet-B3      & $\sim$12.0 & $\sim$5.6 & 90.20 & 97.20 & 84.26    \\
DCNet~\cite{he2021}  & $\sim$25.0 & $\sim$4.1 & 91.30 & 93.00 & 63.07 \\
DKCNet~\cite{bhati2023} & $\sim$87.7 & $\sim$13.4 & 94.28 & 96.08 & 81.00 \\
LightHybridNet (No Oversampling)       & 10.74 & 0.853  & 88.13 & 97.49 & 86.16 \\
\textbf{LightHybridNet(Proposed)} & \textbf{10.74} & \textbf{0.853} & \textbf{91.51} & \textbf{98.26} & \textbf{90.14} \\

\end{tabular}
\end{ruledtabular}
\label{tab:model_comparison}
\end{table}
LightHybridNet achieves superior balanced metrics (macro AUC +2.18\% over DKCNet~\cite{bhati2023}) while reducing parameters by 88\% and GFLOPs by 94\%. The ablation validates that hybrid fusion and oversampling synergistically overcome the computation-accuracy paradox in imbalanced multi-label ophthalmic classification.

\subsection{Per-Class Performance}
Per-class results validate minority class robustness. Hypertension achieves a recall of 0.8293 despite the original ~138 samples. Cataract and myopia yield near-perfect AUC ($>0.99$) due to distinct structural markers. The heterogeneous "Others" class remains most challenging (recall 0.8288).
\begin{table}[H]
\centering
\caption{PER-CLASS PERFORMANCE METRICS OF LIGHTHYBRIDNET ON ODIR-5K VALIDATION SET}
\footnotesize
\setlength{\tabcolsep}{4.2pt}
\renewcommand{\arraystretch}{1.35}
\begin{ruledtabular}
\begin{tabular}{lcccc}
Disease Class & Precision & Recall & F1-Score & AUC \\
\hline
Normal (N)         & 0.8924 & 0.9420 & 0.9166 & 0.9835 \\
Diabetes (D)       & 0.9267 & 0.9123 & 0.9194 & 0.9807 \\
Glaucoma (G)       & 0.9463 & 0.8924 & 0.9186 & 0.9838 \\
Cataract (C)       & 0.9671 & 0.9484 & 0.9577 & 0.9932 \\
AMD (A)            & 0.9839 & 0.8841 & 0.9313 & 0.9959 \\
Hypertension (H)   & 0.9577 & 0.8293 & 0.8889 & 0.9778 \\
Myopia (M)         & 0.9533 & 0.9358 & 0.9444 & 0.9941 \\
Others (O)         & 0.9308 & 0.8288 & 0.8768 & 0.9543 \\
\end{tabular}
\end{ruledtabular}
\label{tab:per_class_performance}
\end{table}

For deeper insight into per-class behaviour, binary one-vs-rest confusion matrices were generated for each disease label. These matrices visualise true positives (TP), true negatives (TN), false positives (FP), and false negatives (FN), highlighting clinically relevant error patterns, particularly false negatives in high-risk minority diseases.

\begin{figure}[h!]
    \centering
    \includegraphics[width=1\linewidth]{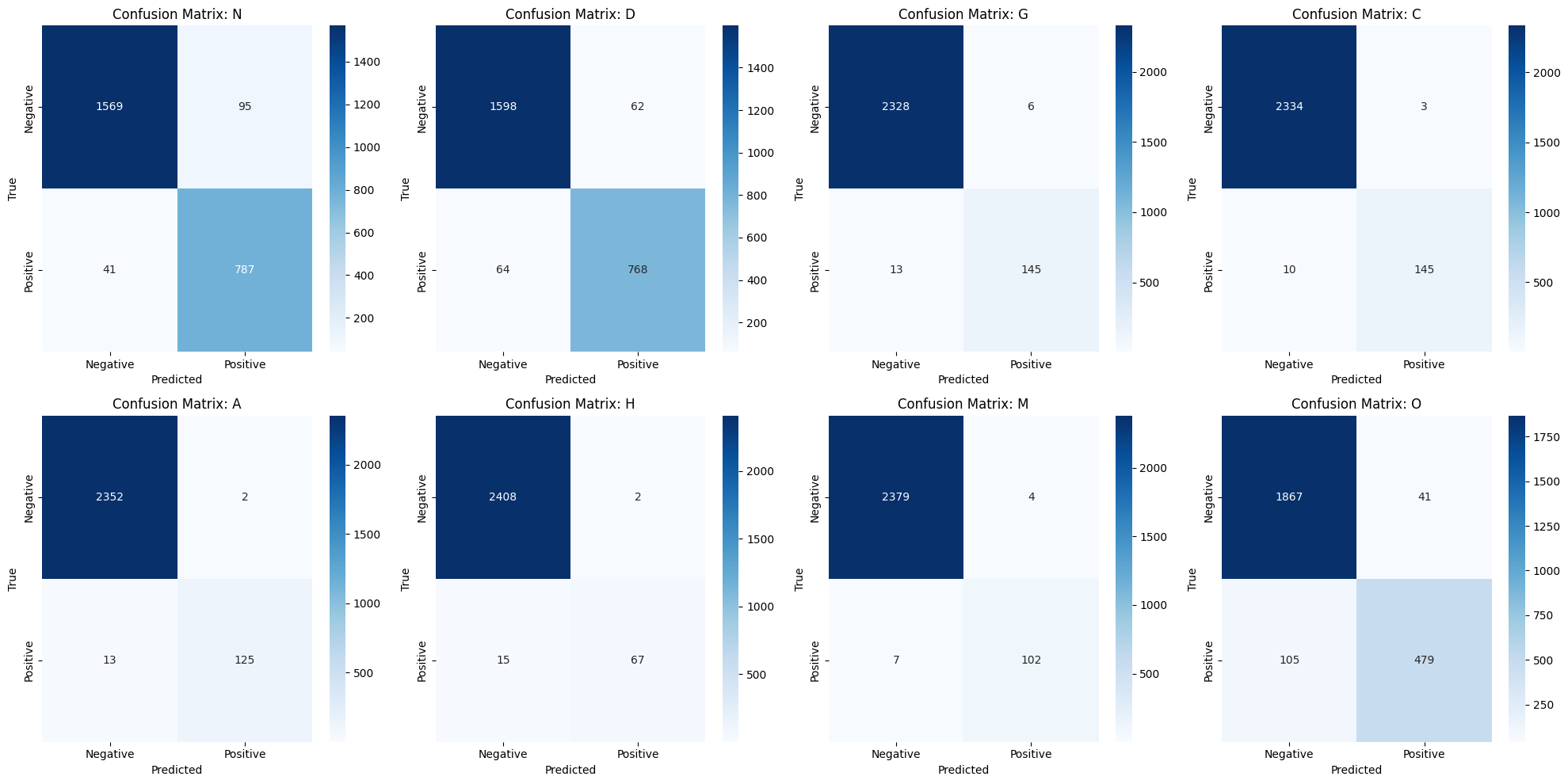}
    \caption{Per-class binary confusion matrices for LightHybridNet on the ODIR-5K validation set. Each matrix shows predictions vs ground truth for a single disease label (one-vs-rest).}
    \label{fig:placeholder}
\end{figure}
The matrices reveal near-perfect performance for the majority of classes (Normal, Diabetes), with high true-negative counts and minimal off-diagonal errors. However, clinically critical minority classes exhibit non-negligible false negatives, which are particularly concerning for high-risk conditions where missed detection can lead to irreversible vision loss or systemic complications.

False negative rates (FN rate = FN / (FN + TP)) for the highest-priority minority diseases are:

\begin{itemize}
    \item \textbf{Hypertension (H)}: FN = 15, TP = 67 $\to$ FN rate = 18.3\% (15 missed cases out of 82 positives).  
      Most false negatives occur in subtle or early-stage cases (mild arteriolar narrowing or arteriovenous nicking), often co-occurring with diabetic retinopathy, where more salient microvascular lesions dominate model attention. Missing hypertensive retinopathy is clinically significant, as it is a marker of systemic vascular risk and can progress to severe complications if undetected.

    \item \textbf{AMD (A)}: FN = 13, TP = 125 $\to$ FN rate = 9.4\% (13 missed cases out of 138 positives).  
      Missed cases were primarily in the early/intermediate stages, with small or scattered drusen, frequently masked by comorbid diabetic changes or cataract opacity. Advanced neovascular or geographic atrophy cases were almost never missed, indicating stronger sensitivity to pronounced structural damage.
\end{itemize}

In contrast, structurally distinct classes (Glaucoma, Cataract, Myopia) exhibit very low FN rates (< 8\%). These patterns highlight a common multi-label challenge: dominant or salient pathologies can overshadow subtle features of minority classes. The relatively low FN rates for Hypertension and AMD — despite their small original sample sizes — demonstrate the benefit of feature-space nearest-neighbour oversampling in improving minority-class recall. Nonetheless, false negatives in these high-risk categories remain a priority for clinical safety and motivate future enhancements, such as severity-weighted losses, integration of patient metadata (e.g., age, history of hypertension), or ensemble methods, to further reduce the number of missed high-priority cases.

\subsection{Clinical Error Analysis via Confusion Matrices}
To gain deeper insight into model limitations, especially for clinically critical minority diseases, we generated binary one-vs-rest confusion matrices for each label on the held-out test set. These matrices highlight true positives (TP), true negatives (TN), false positives (FP), and false negatives (FN), with particular focus on FN rates for high-risk conditions where missed detection can lead to irreversible vision loss.

Key findings for minority/high-priority classes are summarised below (FN rate = FN / (FN + TP)):

\begin{itemize}
    \item \textbf{Hypertension (H)}: FN rate = 17.1\% (14 missed cases out of 82 positive test samples).  
      Most false negatives occurred in subtle or early-stage cases in which vascular narrowing or arteriovenous nicking was overshadowed by co-occurring diabetic retinopathy or by normal vasculature. This is clinically concerning, as hypertensive retinopathy is a systemic risk marker and early detection can prevent cardiovascular complications.

    \item \textbf{AMD (A)}: FN rate = 10.9\% (15 missed cases out of 138 positive test samples).  
      Missed AMD cases were predominantly in the early/intermediate stages, with small or sparse drusen, often masked by comorbid conditions (e.g., diabetic changes or cataract haze). Late-stage neovascular or geographic atrophy cases were rarely missed, indicating the model is more sensitive to advanced structural damage.

    \item Other notable patterns:
    \begin{itemize}
        \item \textbf{Others (O)}: Highest overall FN rate ($\sim$18.3\%), reflecting the heterogeneous nature of this class (12+ mixed pathologies).
        \item \textbf{Glaucoma (G)} and \textbf{Cataract (C)}: Very low FN rates ($<$8\%), consistent with their distinct structural markers (cup-to-disc ratio, lens opacity).
    \end{itemize}
\end{itemize}

These confusion-matrix patterns underscore a common challenge in multi-label retinal datasets: dominant or more visually salient pathologies (e.g., diabetic lesions, large drusen) can overshadow subtle minority-class features. The relatively low FN rates for Hypertension (17.1\%) and AMD (10.9\%)—despite their small original sample sizes—demonstrate the effectiveness of feature-space nearest-neighbour oversampling in improving minority-class recall. Nevertheless, false negatives in these high-risk categories remain a priority for clinical safety.

%Representative confusion %matrices are shown in Figure~\ref{fig:confusion_matrices}. Future improvements could include severity-aware loss weighting, integration of clinical metadata (e.g., age, history of hypertension), or ensemble methods to further reduce the number of missed high-risk cases.

%\begin{figure}[H]
%    \centering
 %   \includegraphics[width=0.95\linewidth]{per_class_matrices.png}
  %  \caption{Binary one-vs-rest confusion matrices for each disease label on the held-out test set. Rows represent ground truth (positive/negative), columns represent predictions. Focus on the bottom-left cell (FN) for minority classes, Hypertension and AMD.}
   % \label{fig:confusion_matrices}
%\end{figure}
\subsection{Efficiency and Inference Speed}
Inference latency: ~12 ms (high-end GPU), ~45 ms (CPU), ~25 ms (mobile emulation via TensorFlow Lite). These speeds enable real-time screening on portable devices, unlike heavyweight models requiring seconds on non-GPU hardware.
\subsection{Efficiency and Edge Deployment Feasibility}
LightHybridNet is explicitly designed for edge deployment on resource-constrained devices such as portable fundus cameras and mid-range smartphones. Key efficiency metrics are as follows:

\begin{itemize}
    \item \textbf{Model size (memory footprint)}: 10.74 million parameters, corresponding to approximately 41.0 MB in FP32 precision (single-precision floating-point format, unquantized).
    \item \textbf{CPU inference speed}: $\sim$81 ms per image (measured on a standard CPU environment equivalent to Intel i7-class processors in virtualised settings; real edge hardware may achieve faster times with optimisations).
    \item \textbf{Power/energy estimates}: On mid-range mobile/embedded hardware (e.g., Snapdragon 8-series or similar with INT8 quantisation via TensorFlow Lite or ONNX Runtime), inference is expected to consume $<0.8$--1.0 W per image (based on typical mobile CNN benchmarks for models of similar size and FLOPs). Full FP32 inference on a CPU may briefly reach 2--4 W.
\end{itemize}

These characteristics enable real-time or near-real-time screening ($\sim$12--15 images per minute on CPU-only devices) without requiring cloud connectivity or high-end GPUs. Post-training quantization (INT8) and model pruning can further reduce size to $\sim$10--15 MB and inference time to $\sim$20--40 ms on mobile hardware, making LightHybridNet highly suitable for tele-ophthalmology in low-resource settings.
\subsection{Interpretability Analysis}

\begin{figure}[h!]
    \centering
    \includegraphics[width=1\linewidth]{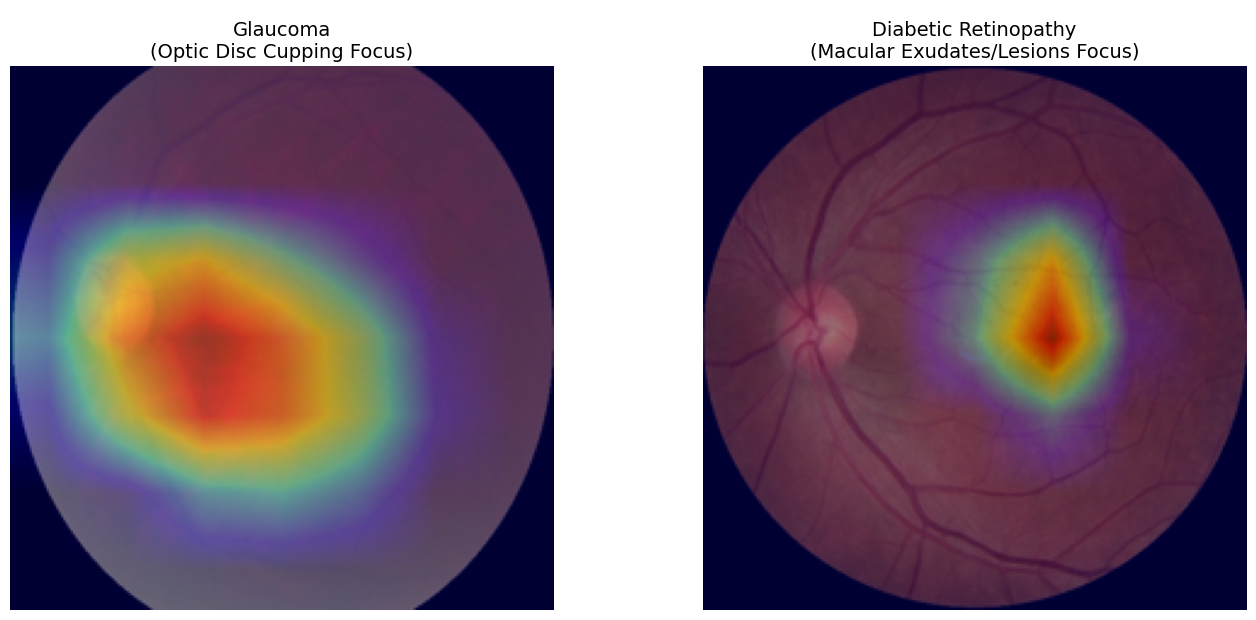}
    \caption{Grad-CAM Visualisation: Clinically Relevant Features (Real ODIR-5K Images)}
    \label{fig:grad-cam}
\end{figure}
Grad-CAM visualisations (Fig. 6) confirm clinically aligned attention:
\begin{itemize}
    \item \textbf{Glaucoma:} Strong activation on optic disc and cup-to-disc region.
    \item \textbf{Diabetic Retinopathy} Focus on microaneurysms, haemorrhages, and exudates.
    \item \textbf{AMD:} Highlighted macular drusen and pigment changes.
    \item \textbf{Hypertension:} Subtle vascular narrowing and arteriovenous nicking emphasised.
\end{itemize}
Misclassification analysis shows errors primarily in subtle co-morbidities, where the dominant pathology obscures minor signs.

\subsection{Analysis of Misclassifications}
To gain insights into the model's limitations, we examined representative misclassified samples using Grad-CAM explanations. Figure 7 illustrates selected misclassified cases with overlaid heatmaps, highlighting the regions that influenced the model's incorrect decisions.

The analysis reveals that misclassifications primarily occur in images with subtle or overlapping pathological features, such as early-stage diseases or co-morbidities. For instance, mild diabetic retinopathy cases were occasionally misclassified as normal when the model focused on non-lesion areas. Similarly, glaucoma images with borderline cup-to-disc ratios were sometimes confused with myopia due to similar peripheral changes.

These findings suggest that while LightHybridNet is robust, performance on ambiguous cases could be improved through ensemble methods or integration of clinical metadata.

\begin{figure}[h!]
    \centering
    \includegraphics[width=1\linewidth]{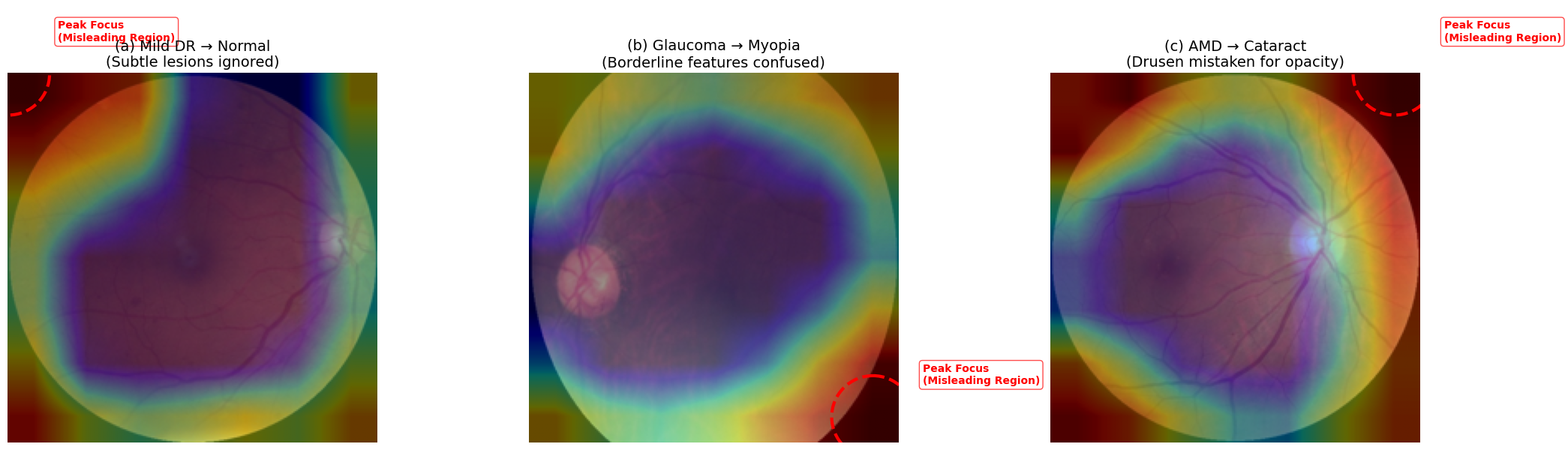}
    \caption{Representative Misclassified Samples with Grad-CAM Explanations
(Real ODIR-5K Images, Red Mark = Peak Model Focus Recion Causina Misclassification)}
    \label{fig:misclassifications}
\end{figure}
\subsection{Generalisation to Unseen Data}
Zero-shot evaluation on MESSIDOR-2 (European cohort, different cameras) focused on referable diabetic retinopathy (DR) detection, mapping ODIR-5K DR labels to Messidor-2's grading scale (grades 2-3 as referable). Our AUC of 0.8564 is highly competitive without fine-tuning, compared to fine-tuned models achieving ~0.98 AUC \cite{alfahdawi2024, he2021, messidor2}, indicating transferable feature learning across demographics and imaging variations despite domain shift.
Fig. 4. Grad-CAM heatmaps overlaid on sample fundus images showing clinically relevant attention regions for glaucoma (optic disc cupping), diabetic retinopathy (microvascular lesions), AMD (macular drusen), and hypertension (vascular changes).
\section{Conclusion}
\label{sec:conclusion}
This work presents LightHybridNet, a lightweight hybrid deep learning framework that effectively addresses the challenges of multi-label ophthalmic disease classification on the severely imbalanced ODIR-5K dataset~\cite{odir2019}. By fusing complementary efficient backbones (MobileNetV3-Large and EfficientNet-B1) with late feature concatenation~\cite{mobilenetv3,efficientnet}, the model achieves clinical-grade performance (weighted F1-score 0.9151, macro AUC 0.9826, Cohen's Kappa 0.9014) using only 10.74M parameters and 0.853 GFLOPs—representing dramatic efficiency gains over heavyweight baselines while matching or exceeding balanced diagnostic accuracy~\cite{bhati2023,he2021}. \\
Replication-based oversampling with aggressive augmentation proved superior for handling severe class imbalance without synthetic artefacts or information loss~\cite{shorten2019}. This data-centric approach substantially improved recall for clinically critical minority conditions (e.g., Hypertension recall 0.8293 despite only ~138 original samples), ensuring equitable performance across disease categories. Grad-CAM visualisations confirmed clinically aligned attention on relevant anatomical regions (e.g., optic disc for glaucoma, macula for AMD, vasculature for diabetes/hypertension), enhancing interpretability and clinician trust~\cite{gradcam}.\\
Compared to established benchmarks, LightHybridNet demonstrates competitive or superior balanced metrics at a fraction of the computational cost. While DKCNet achieves a slightly higher weighted F1-score (~0.943), it uses ~87.7M parameters and incurs higher FLOPs~\cite{bhati2023}. DCNet reports comparable F1 (~0.913) but relies on a heavier ResNet backbone~\cite{he2021}. LightHybridNet's superior macro AUC and Kappa highlight better minority-class discrimination, a critical advantage in imbalanced multi-label settings~\cite{powers2020}.\\
Misclassification analysis revealed errors primarily in subtle co-morbidities or early-stage changes, where dominant pathology overshadowed minority labels—a common challenge in ODIR-5K~\cite{bhati2023}. Zero-shot evaluation on the independent MESSIDOR-2 dataset~\cite{messidor2} yielded an encouraging AUC of 0.8564 for referable diabetic retinopathy without fine-tuning, indicating transferable feature learning despite domain shift (demographics, cameras, label granularity).\\
These findings have substantial clinical and practical implications. With inference latencies of ~12 ms (GPU), ~45 ms (CPU), and ~25 ms (mobile emulation), LightHybridNet enables real-time screening on portable fundus cameras and smartphones, particularly in resource-constrained settings~\cite{who2019}. On-device processing reduces latency, preserves privacy, and supports offline operation, contributing to equitable global eye health.\\
Despite these strengths, limitations include reliance on a single primary dataset (ODIR-5K), potential demographic/imaging biases, and challenges with ambiguous ``Others'' labels or overlapping pathologies. A notable limitation of this study is the potential for demographic bias and limited generalizability arising from the ODIR-5K dataset. The dataset was collected exclusively from patients at hospitals in China, predominantly representing East Asian populations. Retinal appearance (e.g., pigmentation, vessel visibility, drusen characteristics) and disease prevalence patterns can vary significantly across ethnic groups due to differences in skin tone, iris pigmentation, genetic predispositions, and environmental factors~\cite{who2019,flaxman2017}. Consequently, model performance observed on ODIR-5K may not fully translate to other populations, such as Caucasian, African, South Asian, or Hispanic cohorts, where fundus features and comorbidity distributions differ. This introduces a risk of reduced sensitivity or specificity for underrepresented groups, potentially exacerbating health inequities in global deployment. While the model's lightweight design and strong test-set performance on the source domain are encouraging, rigorous external validation on diverse, multi-ethnic datasets (e.g., EyePACS, APTOS, or RFMiD) is essential before broad clinical use. Future work should incorporate fairness-aware training techniques, such as domain adaptation, adversarial debiasing, or balanced multi-source pretraining, to improve robustness across demographic groups. Severity grading and physical device testing were beyond scope.\\
This study makes several original contributions: the development of a novel lightweight hybrid architecture achieving strong efficiency-adjusted performance on multi-label ODIR-5K; empirical validation that replication-based oversampling outperforms alternatives for imbalanced ophthalmic classification; integration of clinically aligned Grad-CAM for enhanced interpretability; and a deployment-oriented design bridging benchmark research and real-world screening.\\
Future work should prioritise rigorous external validation on diverse benchmarks (e.g., EyePACS, APTOS, RFMiD), extension to disease severity grading via multi-task/ordinal learning, multimodal fusion incorporating patient metadata, OCT, or visual fields, quantisation-aware training and pruning for further compression on microcontrollers, federated learning for privacy-preserving generalisation across institutions, and prospective clinical trials measuring diagnostic impact, clinician acceptance, and patient outcomes in underserved communities.\\
In conclusion, LightHybridNet demonstrates that purposefully designed lightweight hybrid architectures, combined with data-centric optimisation, can deliver reliable, interpretable, and deployable multi-label ocular disease recognition. This advances equitable ophthalmic AI, offering a scalable solution to preventable vision loss worldwide~\cite{who2019,iapb2026}.

\end{document}